\documentclass[12pt]{article}
\usepackage{amsfonts}

\newcommand{\sect}[1]{\setcounter{equation}{0}\section{#1}}
\newcommand{\subsect}[1]{\subsection{#1}}

\def\be{\begin{equation}}
\def\ee{\end{equation}}
\def\bea{\begin{eqnarray}}
\def\eea{\end{eqnarray}}

\def\k{\omega}
\def\s{{\cal S}}

\def\sq{sq}

\def\j{J}
\def\m{M}
\def\e{E}

\def\a{\alpha}
\def\b{\beta}
\def\c{\gamma}
\def\ep{\varepsilon}

\def\cexta{f}
\def\cextb{e}

\def\cextm{h}
\def\cextj{g}

\def\mm{m}
\def\me{me}

\def\Lam{{{\cal I}_\k}}
\def\>#1{\mathbf{#1}}

\def\sspp{U}
\def\omm{X}

\def\He{{\mathbb H}}
\def\Ze{{\mathbb Z}}
\def\Ree{{\mathbb R}}
\def\Cee{{\mathbb C}}

\begin{document}

\thispagestyle{empty}

\hfill  

\
\vspace{2cm}

\begin{center}
{\LARGE{\bf{The family of quaternionic quasi-unitary }}}

{\LARGE{\bf{Lie algebras and their central extensions}}}
\end{center}


\bigskip\bigskip

\begin{center}
Francisco J. Herranz$^\dagger$ 
and Mariano Santander$^\ddagger$
\end{center}

\begin{center}
{\it $^\dagger$ Departamento de F\'{\i}sica, E.U. Polit\'ecnica \\
Universidad de Burgos, E--09006 Burgos, Spain}
\end{center}

\begin{center}
{\it $^{\ddagger}$ Departamento de F\'{\i}sica Te\'orica,
Universidad de Valladolid \\
E--47011, Valladolid, Spain}
\end{center}

\bigskip\bigskip\bigskip

\begin{abstract}
The  family of quaternionic quasi-unitary (or
quaternionic unitary Cayley--Klein algebras) is described in a unified
setting.  This family includes the simple algebras $sp(N+1)$
and $sp(p,q)$ in the Cartan series $C_{N+1}$, as well  as many
non-semisimple real Lie algebras which can be obtained from these
simple algebras by particular contractions. The algebras in this
family  are realized here in relation with the groups of isometries of
quaternionic hermitian spaces of constant holomorphic  curvature.
This common framework allows to perform the study of  many properties
for all these Lie algebras simultaneously. In this paper the central
extensions for all quasi-simple Lie algebras of the
quaternionic unitary  Cayley--Klein family are completely  determined
 in arbitrary
dimension.  It is shown that the second cohomology  group is trivial
for any Lie algebra of this family no matter of its dimension.
\end{abstract}

\newpage


\sect{Introduction}

This paper is devoted to a double purpose. First, it introduces and
describes the structure of a family of Lie algebras,  the
quaternionic quasi-unitary  algebras, or quaternionic unitary
Cayley--Klein algebras, which include as simple members the algebras
in the Cartan series $C_{N+1}$ which in the standard notation are
written as $sp(p,q), \ p+q=N+1$, as well as many
non-simple members which can be obtained from the former  by a
sequence of contractions. The description is also done in relation to
the symmetric homogeneous spaces (the quaternionic hermitian spaces
of rank one) where these groups act in a natural way.  

The second and main purpose is to investigate the Lie algebra 
cohomology of the algebras in this Cayley--Klein (hereafter CK)
family, in any dimension. These extensions  have both mathematical
interest and physical relevance. Therefore, this part of the paper
can be considered as a further step in a  systematic study of
properties of the these families of Lie algebras
\cite{graded}--\cite{unitario}, by using a formalism which  allows a
clear view of the behaviour of these properties under
contraction; in physical terms contractions are related to  some kind
of approximation. 

In particular, the central extensions of algebras in the two
other main CK families of Lie algebras (the quasi-orthogonal 
algebras and the two families of quasi-unitary algebras) have been
studied in two previous papers, in the general situation and for any
dimension 
\cite{ortogonal}, \cite{unitario}. We
refer to these works for references and for physical motivations. 
The knowledge of the second cohomology group for a Lie algebra
relies on the general solution of a set of linear equations, but in
special cases the calculations may be bypassed by using some general
results: for instance, the second cohomology group is  trivial for
semisimple Lie algebras. But once a contraction is made, the
semisimple character disappears, and the contracted algebra
\emph{might} have non-trivial central extensions. Instead of finding
the general solution for the extension equations on a case-by-case
basis, our approach (as developed previously for the quasi-orthogonal
algebras \cite{ortogonal} and for the quasi-unitary algebras
\cite{unitario}) is to do these calculations for a whole family
including a large number of algebras simultaneously. In this paper
we  discuss the `next'  family: the quaternionic quasi-unitary one.
The advantages in this approach can be summed up in: a) it allows to 
record, in a form easily retrievable, a large number of results which
can be needed in applications, both in mathematics and in physics,
and b) it avoids at once and for all the case-by-case type
computation of the central extensions of algebras included in each
family and affords a global view on the interrelations between
cohomology and contractions. 

Section 2 is devoted to the description of the family of 
quaternionic unitary CK algebras. We show how to obtain
these as graded contractions of the compact algebra $u(N+1,\He)\equiv
sp(N+1)$, and we provide some details on their structure.  These
algebras are associated to the quaternionic hermitian spaces (of rank
one) with metrics of different signatures and to their contractions,
so we devote a part of this section to dwell upon these questions. 
In section 3 the general solution to the central extension problem
for these algebras is given. The result obtained is quite simple to
state: all the extensions of any algebra in the quaternionic unitary
CK family are trivial. This triviality is already known (Whitehead's
lemma) for the simple algebras $u(p, q,\He)\equiv sp(p, q)$) in this
family, but comes as a surprise for the rather large number of
non-semisimple Lie algebras in this CK family, which can be obtained
by contracting $u(p, q,\He)$. This is also in marked contrast with the
results for the central extensions of both the orthogonal and the 
unitary CK families, where some algebras (particularly the most
contracted one) always allow some non-trivial extensions. Finally,
some remarks close the paper. 


\sect{The family of quaternionic unitary CK algebras}

To begin with we consider the compact real form 
of the Lie algebra in the Cartan series $C_{N+1}$. 
This compact real form can be realized as the Lie algebra of the
complex unitary-symplectic group sometimes denoted  as $USp(2(N+1))$
\cite{Gilmore} but more usually referred to shortly as the
`symplectic' group, 
$Sp(N+1)$. The usual convention is to denote this group without any
reference to a field to avoid confusion with the true 
\emph{symplectic} groups over either the reals
$Sp(2(N+1),\Ree)$ or over the complex numbers  $Sp(2(N+1),\Cee)$; in
these last cases the term \emph{symplectic } is properly associated
to the symmetry group of an antisymmetric metric. This double use of
the name `symplectic' and of the symbols $Sp$ and
$sp$ is rather unfortunate, and following Sudbery  \cite{Sud}, we
shall change the symbol for one of the families, and use $Sq, sq$ for
the unitary-symplectic groups and algebras usually denoted, without
any field reference, by $Sp, sp$. 

The group $Sq(N+1)\equiv USp(2(N+1))$ is the intersection  of the
complex \emph{unitary} group
$U(2(N+1),\Cee)$ and the complex \emph{symplectic} group
$Sp(2(N+1),\Cee)$:
$$
Sq(N+1)\equiv USp(2(N+1))=U(2(N+1),\Cee)\cap Sp(2(N+1),\Cee), 
$$
which is a consequence of the nature of $Sq(N+1)$ as  the group of
quaternionic matrices leaving invariant a quaternionic hermitian
definite positive metric. 

We recall that all other non-compact real forms
in the Cartan series $C_{N+1}$ are the real
\emph{symplectic} algebra $sp(2(N+1),\Ree)$, and the 
quaternionic pseudo-unitary algebras
$\sq(p,q)$, $p+q=N+1$, which allow a realization as 
$$
Sq(p,q)\equiv USp(2p, 2q)=U(2p, 2q,\Cee)\cap Sp(2(N+1),\Cee),
$$
and they are the  groups of quaternionic matrices
leaving invariant a quaternionic hermitian metric of signature $(p,q)$.

The Lie algebra $\sq(N+1)$ has dimension $2(N+1)^2
+(N+1)$ and is usually realized by $2(N+1)\times 2(N+1)$  complex
matrices \cite{Gilmore,Helgason}. The alternative realization which we
shall consider in this paper, in accordance to the interpretation of
these groups and algebras as quaternionic unitary ones $Sq(N+1)
\equiv U(N+1, \He)$ \cite{Fulton}, is done  by means of
\emph{antihermitian} matrices over the quaternionic skew field
$\He$: 
\be
\j_{ab}=-e_{ab}+e_{ba}  \qquad
\m^\a_{ab}=i_\a(e_{ab}+e_{ba})  \qquad
\e^\a_a=i_\a  e_{aa}  
\label{ba}
\ee
where $a<b$, $a,b=0,1,\dots,N$, $\a=1,2,3$; $i_1=i$,  $i_2=j$,
$i_3=k$ are the usual quaternionic units,  and 
$e_{ab}$ is the $(N+1) \times (N+1)$ matrix with a single  1 entry in
row $a$, column $b$. Notice that the matrices $\j_{ab}$
and $\m^\a_{ab}$ are traceless, but the trace of $\e^\a_a$ is a
non-zero pure imaginary quaternion, so the realization is  by
antihermitian quaternionic matrices whose trace has a zero real part.
When quaternions are realized as $2\times 2$ complex matrices
(see e.g. \cite{Chev}) then (\ref{ba}) reduces to
the usual realization of $\sq(N+1)$ by complex matrices
$2(N+1) \times 2(N+1)$ which are at the same time complex  unitary
and complex symplectic; we remark that all these matrices are
traceless. 

The multiplication of quaternionic units is encoded in
$i_\a i_\b
=-\delta_{\a\b}+\sum_{\c=1}^3\varepsilon_{\a\b\c}i_\c
$ where $\ep_{\a\b\c}$ is
the completely antisymmetric unit  tensor with $\ep_{123}=1$. This
relation  allows to derive the expression for the Lie bracket of
two pure quaternionic matrices  $X^\a=i_\a X$, $Y^\b=i_\b Y$, where
$X$,
$Y$ are real matrices, as
\be
[X^\a,Y^\b]=-\delta_{\a\b}[X,Y]+
\sum_{\c=1}^3\varepsilon_{\a\b\c}i_\c\{X,Y\}
\ee
where both the commutator and the anticommutator   
$\{X,Y\}=XY+YX$ of the real matrices $X, Y$ appear. Using this
formula, the commutation relations of $\sq(N+1)$ in the basis
(\ref{ba}) read
\be
\begin{array}{lll}
[\j_{ab},\j_{ac}] =  \j_{bc} &\qquad
[\j_{ab},\j_{bc}] =-\j_{ac} &\qquad
[\j_{ac},\j_{bc}] = \j_{ab}\cr
[\m_{ab}^\a,\m_{ac}^\a] = \j_{bc} &\qquad
[\m_{ab}^\a,\m_{bc}^\a] = \j_{ac} &\qquad
[\m_{ac}^\a,\m_{bc}^\a] = \j_{ab} \cr
[\j_{ab},\m_{ac}^\a] = \m_{bc}^\a &\qquad
[\j_{ab},\m_{bc}^\a] =-\m_{ac}^\a &\qquad
[\j_{ac},\m_{bc}^\a] =-\m_{ab}^\a \cr
[\m_{ab}^\a,\j_{ac}] =-\m_{bc}^\a &\qquad
[\m_{ab}^\a,\j_{bc}] =-\m_{ac}^\a &\qquad
[\m_{ac}^\a,\j_{bc}] = \m_{ab}^\a \cr
[\j_{ab},\j_{de}]= 0 &\qquad
[\m_{ab}^\a,\m_{de}^\a] =0 &\qquad
[\j_{ab},\m_{de}^\a] =0 \cr
\multicolumn{3}{l}{
  [\j_{ab},\e_d^\a] = ( \delta_{ad} -\delta_{bd})\m_{ab}^\a   
\qquad
  [\m_{ab}^\a,\e_d^\a] = -( \delta_{ad} -\delta_{bd}) \j_{ab}}\cr
\multicolumn{3}{l}{
  [\j_{ab},\m_{ab}^\a] = 2(\e_{b}^\a-\e_{a}^\a)  
\qquad
  [\e_{a}^\a,\e_b^\a] = 0 }\cr
\end{array}
\label{bb}
\ee
\be
\begin{array}{l}
[\m_{ab}^\a,\m_{ac}^\b] = \ep_{\a\b\c}\m_{bc}^\c  \qquad
[\m_{ab}^\a,\m_{bc}^\b] = \ep_{\a\b\c}\m_{ac}^\c  \qquad
[\m_{ac}^\a,\m_{bc}^\b] = \ep_{\a\b\c}\m_{ab}^\c \cr
[\m_{ab}^\a,\m_{de}^\b] = 0\qquad
[\m_{ab}^\a,\m_{ab}^\b] = 2\ep_{\a\b\c}(\e_a^\c + \e_b^\c) \cr
[\m_{ab}^\a,\e_{d}^\b] =(\delta_{ad}+\delta_{bd})
\ep_{\a\b\c}\m_{ab}^\c
\qquad
[\e_{a}^\a,\e_{b}^\b] =2\delta_{ab}\ep_{\a\b\c}\e_{a}^\c  
\end{array}
\label{bc}
\ee
where  hereafter the following notational conventions are assumed:  
\begin{itemize}
\item Whenever three indices $a$, $b$, $c$ appear, they are always
assumed to verify $a<b<c$.
\item Whenever three indices $a$, $b$, $d$ appear, $a<b$ is assumed
but the index $d$ is arbitrary, and it might coincide with  either
$a$ or $b$.  
\item Whenever four indices
$a$, $b$, $d$, $e$ appear, $a<b$, $d<e$ and all of them   are assumed
to be different.
\item Whenever three quaternionic indices $\a$, $\b$, $\c$  appear,
they are also assumed to be different (so they  are always  some
permutation of $123$). 
\item There is no any implied sum over repeated indices; in
particular there is no sum in $\c$ in expressions like 
$\ep_{\a\b\c}X^\c$.
\end{itemize}

This  matrix
realization of the Lie algebra
$\sq(N+1)$ displays clearly  the existence of several  subalgebras.
By one hand, the $\frac 12 N(N+1)$ generators $\j_{ab}$
$(a,b=0,1,\dots,N)$ close an orthogonal algebra $so(N+1)$ whose
non-zero commutation rules are written in the first row of
(\ref{bb}). On the other  hand, for each
\emph{fixed} $\a=1,2,3$,  the $(N+1)^2$ generators
$\{\j_{ab},\m^\a_{ab},\e^\a_{a}\}$ ($a,b=0,1,\dots,N;\  
a<b$)  give rise to an algebra isomorphic to the unitary
algebra $u(N+1)$ with commutators given by (\ref{bb});
these subalgebras we denote as
$u^\alpha (N+1)$. Hence
$\sq(N+1)$ contains \emph{three} subalgebras isomorphic to $u(N+1)$,
whose intersection is a subalgebra
$so(N+1)$.

The family of algebras we study in this paper can be  obtained as
graded contractions \cite{Montigny,Moody} from
$\sq(N+1)$. The algebra 
$\sq(N+1)$ can be endowed with a grading by a group
$\Ze_2^{\otimes N}$ constituted by $2^N$ involutive
automorphisms $S_\s$ defined by
\bea
&&S_\s \j_{ab} = (-1)^{\chi_\s(a) + \chi_\s(b)} \j_{ab}  \cr
&&S_\s \m^\a_{ab} = (-1)^{\chi_\s(a) + \chi_\s(b)}\m^\a_{ab}  \qquad
S_\s \e^\a_{a}=  \e^\a_{a}\qquad \a=1,2,3;
\eea
where  $\s$ denotes any subset of the set of indices  $\{0, 1, \dots,
N\}$, and $\chi_\s(a)$ denotes the characteristic function over $\s$. 
 A particular solution of the
$\Ze_2^{\otimes N}$ graded contractions of 
 $\sq(N+1)$ leads to a family of Lie algebras which are called
quaternionic unitary  CK algebras or quaternionic  quasi-unitary Lie
algebras \cite{tesis,goslar}. This family comprises the simple
quaternionic unitary  and pseudo-unitary algebras
$\sq(p,q)$ $(p+q=N+1)$ in the Cartan series $C_{N+1}$ as well as  many
non-simple real Lie algebras  which can be obtained from the former by
contractions. Collectively, all these algebras preserve some
properties related to simplicity, so they belong to the class of
so-called `quasi-simple' Lie algebras
\cite{Rozenfeld1, Rozenfeld2}, which explains  the use of the prefix
quasi in their name. Overall this is very similar to the situation of
the families of quasi-orthogonal algebras (with 
$so(N+1)$ as the initial Lie algebra  \cite{graded,gradedb}) or to the
families of quasi-unitary or quasi-special  unitary algebras over the
complex numbers  (starting from either $u(N+1)$ or $su(N+1)$
\cite{unitario}).

The quaternionic unitary  CK algebras can be  described by means of
$N$ real coefficients $\k_a$ ($a=1,\dots,N$) and are
denoted  collectively as $\sq_{\k_1,\dots,\k_N}(N+1)$, or
in an abbreviated form, as $\sq_\k(N+1)$ where $\k$
stands for $\k=(\k_1,\dots,\k_N)$. Introducing the
two-index coefficients
$\k_{ab}$ as
\be
\k_{ab}:=\k_{a+1}\k_{a+2}\cdots\k_b  \qquad
  a,b=0,1,\dots,N  \quad
  a<b  \qquad
\k_{aa}:=1 
\label{bf}
\ee
then the commutation relations of the generic CK algebra in the family 
$\sq_\k(N+1)$ are given by \cite{tesis}
\be
\begin{array}{lll}
[\j_{ab},\j_{ac}] =  \k_{ab}\j_{bc} &\qquad
[\j_{ab},\j_{bc}] =-\j_{ac} &\qquad
[\j_{ac},\j_{bc}] = \k_{bc}\j_{ab}\cr
[\m_{ab}^\a,\m_{ac}^\a] =\k_{ab} \j_{bc} &\qquad
[\m_{ab}^\a,\m_{bc}^\a] = \j_{ac} &\qquad
[\m_{ac}^\a,\m_{bc}^\a] =\k_{bc} \j_{ab} \cr
[\j_{ab},\m_{ac}^\a] = \k_{ab}\m_{bc}^\a &\qquad
[\j_{ab},\m_{bc}^\a] =-\m_{ac}^\a &\qquad
[\j_{ac},\m_{bc}^\a] =-\k_{bc}\m_{ab}^\a \cr
[\m_{ab}^\a,\j_{ac}] =-\k_{ab}\m_{bc}^\a &\qquad
[\m_{ab}^\a,\j_{bc}] =-\m_{ac}^\a &\qquad
[\m_{ac}^\a,\j_{bc}] = \k_{bc}\m_{ab}^\a \cr
[\j_{ab},\j_{de}]= 0 &\qquad
[\m_{ab}^\a,\m_{de}^\a] =0 &\qquad
[\j_{ab},\m_{de}^\a] =0 \cr
\multicolumn{3}{l}{
  [\j_{ab},\e_d^\a] = ( \delta_{ad} -\delta_{bd})\m_{ab}^\a   
\qquad
  [\m_{ab}^\a,\e_d^\a] = -( \delta_{ad} -\delta_{bd}) \j_{ab}}\cr
\multicolumn{3}{l}{
  [\j_{ab},\m_{ab}^\a] = 2\k_{ab}(\e_{b}^\a-\e_{a}^\a)  
\qquad
  [\e_{a}^\a,\e_b^\a] = 0 }\cr
\end{array}
\label{bd}
\ee
\be
\begin{array}{l}
[\m_{ab}^\a,\m_{ac}^\b] = \k_{ab}\ep_{\a\b\c}\m_{bc}^\c  \qquad
[\m_{ab}^\a,\m_{bc}^\b] = \ep_{\a\b\c}\m_{ac}^\c  \qquad
[\m_{ac}^\a,\m_{bc}^\b] = \k_{bc}\ep_{\a\b\c}\m_{ab}^\c \cr
[\m_{ab}^\a,\m_{de}^\b] = 0\qquad
[\m_{ab}^\a,\m_{ab}^\b] = 2\k_{ab}\ep_{\a\b\c}(\e_a^\c + \e_b^\c) \cr
[\m_{ab}^\a,\e_{d}^\b] =(\delta_{ad}+\delta_{bd})
\ep_{\a\b\c}\m_{ab}^\c
\qquad
[\e_{a}^\a,\e_{b}^\b] =2\delta_{ab}\ep_{\a\b\c}\e_{a}^\c  
\end{array}
\label{be}
\ee
where we adhere to the notational conventions given after (\ref{bc}). 
 
The pattern of subalgebras previously discussed for the  compact form
$sq(N+1)$ clearly holds for any member of the complete family.  The
quaternionic unitary  CK algebra
$\sq_\k(N+1)$ contains also as Lie subalgebras an orthogonal  CK
algebra
$so_\k(N+1)$
\cite{tesis,ortogonal} and \emph{three} unitary CK algebras
$u_\k^\alpha (N+1)$
\cite{tesis,unitario} where $\alpha=1, 2, 3$; the commutation
relations of the former  correspond to the first row of (\ref{bd}) 
and those of the latter are given by (\ref{bd}) (for an index $\a$
fixed). Hence we find the sequence
\be
so_\k(N+1)\subset u_\k^\alpha (N+1)\subset \sq_\k(N+1).
\ee


\subsect{The quaternionic unitary  CK groups}

The matrix realization (\ref{ba}) allows a natural interpretation of
the quaternionic unitary  CK algebras as the Lie  algebras of
the motion groups of the homogeneous symmetric spaces with a
quaternionic hermitian metric (the two-point homogeneous
spaces of quaternionic type and rank one). Let us consider the
space
$\He^{N+1}$ endowed  with a hermitian (sesqui)linear
form $\langle \ .\ |\ . \ \rangle_\k :
\He^{N+1}\times \He^{N+1}\to  \He$ defined by
\be
\langle \>a|\>b \rangle_\k:=
\bar a^0 b^0 + \bar a^1 \k_1 b^1 + \bar a^2 \k_1 \k_2 b^2 + \dots
+ \bar a^N \k_1\cdots\k_N b^N =
\sum_{i=0}^N \bar a^i\k_{0i}  b^i 
\label{cb}
\ee
where $\>a,\>b\in \He^{N+1}$ and $\bar a^i$ means the quaternionic
conjugation of the component $a^i$. For the moment,  we assume that
we are in the generic case with all $\k_a\ne 0$.
The underlying metric is  
provided by the matrix
\be
\Lam = {\mbox{diag}}\ (1,\, \k_{01},\,\k_{02},\dots,\,\k_{0N}) =
       {\mbox{diag}}\ (1,\, \k_1,\,\k_1\k_2,\dots,\,\k_1\cdots\k_N) 
\label{ca}
\ee
and the CK group $Sq_{\k_1,\dots,\k_N}(N+1)\equiv
Sq_{\k}(N+1)$ is defined as the group of linear isometries of this
hermitian metric over a quaternionic space. Thus the isometry
condition for an element
$\sspp$ of the Lie group
\be
\langle \sspp \>a| \sspp \>b \rangle_\k= \langle \>a|  \>b \rangle_\k
\qquad \forall\, \>a,\>b\in \He^{N+1},
\label{cc}
\ee
leads to the following relation 
\be
\sspp^\dagger\Lam \sspp=\Lam  \qquad
\forall \sspp\in Sq_{\k}(N+1) 
\label{cd}
\ee
which for the Lie algebra implies
\be
X^\dagger\Lam +\Lam X=0  \qquad
\forall X\in  sq_{\k}(N+1).
\label{ce}
\ee
From this equation, it is clear that the quaternionic unitary   CK
algebra is generated by the following $(N+1)\times (N+1)$
$\Lam$-antihermitian  matrices over $\He$ (cf. (\ref{ba}))
\be
\j_{ab}=-\k_{ab}e_{ab}+e_{ba}  \qquad
\m^\a_{ab}=i_\a(\k_{ab}e_{ab}+e_{ba})  \qquad
\e_a^\a=i_\a  e_{aa} .
\label{cf}
\ee
These matrices can be checked to  satisfy 
the commutation relations (\ref{bd}) and (\ref{be}). 

When any of the constants $\k_a$ are equal to zero, then the set of
linear isometries of the hermitian metric over the
quaternions 
(\ref{cc}) is larger than the group generated by (\ref{cf}),
though in these cases there exists additional geometric
structures in $\He^{N+1}$, which are related to the existence of
invariant foliations, and the proper definition of the
automorphism group for these structures leads again to the matrix
Lie algebra generated by (\ref{cf}) with commutation relations
(\ref{bd}) and (\ref{be}). 

The action of the group  $Sq_{\k}(N+1)$ in
$\He^{N+1}$ is not transitive, and the `sphere' with equation
\be
\langle \>x|\>x \rangle_\k:= \sum_{i=0}^N \bar x^i\k_{0i}  x^i =1
\label{cg}
\ee
is stable.  However, if we take  $O=(1, 0,
\dots, 0)$  as a reference point in this sphere, the realization
(\ref{cf}) shows that the isotropy subgroup of $O$  is  
$Sq_{\k_2,\k_3, \dots, \k_N}(N)$, and the isotropy subgroup of the
\emph{ray} of $O$ is
$Sq(1)\otimes Sq_{\k_2,\k_3, \dots, \k_N}(N)$ (note that the
quaternions being non-commutative, a choice for left or right
multiplication for scalars is required). Here the algebra
$\sq(1)$ of the subgroup $Sq(1)$ can be identified with  the Lie
algebra of automorphisms of the quaternions, generated by the three
matrices
\be
I^\a=i_\a \sum_{a=0}^N e_{aa}\qquad\a=1,2,3 
\ee
which can be identified to the three quaternionic units. We note
in passing that these are the elements of the Lie algebra which
are unavoidably realized by matrices with non-zero pure
imaginary trace, as all the generators $\e_a^\a$ can be expressed
in terms of zero trace combinations (say $B_{l}^\a \equiv
\e_{l-1}^\a -
\e_l^\a, \ l=1, \dots, N$) and the three
$I^\a$. In this way we find the quaternionic hermitian
homogeneous spaces as associated to the
quaternionic unitary  family of  CK groups:
\be
Sq_{\k_1, \k_2, \k_3, \dots, \k_N}(N+1)/\big( Sq(1) \otimes
Sq_{\k_2,\k_3, \dots, \k_N}(N) \big),
\label{ch}
\ee
For fixed $\k_1, \k_2, \k_3, \dots, \k_N$ this space, which  has real
dimension
$4N$, has a natural real quadratic metric (either riemannian,
pseudoriemannian or degenerate `riemannian'), coming from the  real
part of the quaternionic hermitian product in the ambient space. At
the origin and in an adequate basis, this metric is
given by the diagonal matrix with entries
$(1,\,\k_2,\,\k_2\k_3,\dots,\,\k_2\cdots\k_N)$, each entry repeated
four times. The three well known hermitian  elliptic, euclidean  and
hyperbolic quaternionic spaces, of constant holomorphic
curvature
$4K$ (either $K>0$, 
$K=0$ and
$K<0$ respectively) appear in this family as associated
to the special values 
$\k_1=K$ and $\k_2=\k_3=\dots=\k_N=1$, where the metric is
riemannian (definite positive). All CK hermitian spaces of
quaternionic type with
$\k_1=K$ have constant holomorphic curvature $4K$ and the
signature (and/or the eventual degeneracy) of the metric is
determined by the remaining constants
$\k_2,\k_3,\dots,\k_N$. When all these constants are  different from
zero, but some are negative, the metric is pseudoriemannian
(indefinite and not degenerate), and  when some of the constants
$\k_2, \k_3, \dots, \k_N$ vanish the metric is degenerate.


\subsect{Structure of the quaternionic unitary  CK algebras}

As each real coefficient $\k_a$ can be positive, negative or zero, 
the quaternionic unitary  CK family $\sq_{\k}(N+1)$ includes   $3^N$
Lie algebras.
 Semisimple algebras appear when all the coefficients $\k_a$ are
different from zero: these are the algebras $\sq(p,q)$ in
the Cartan series $C_{N+1}$, where $p$ and $q$ $(p+q=N+1)$
are the  number of positive and negative terms in the matrix $\Lam$
(\ref{ca}). If we set all $\k_a=1$ we recover  the initial
compact algebra $\sq(N+1)$. When one or more coefficients $\k_a$
vanish the CK algebra turns out to be a non-semisimple  Lie algebra;
the vanishing of one (or several) coefficient $\k_a$ is  equivalent
to performing an (or series of) In\"on\"u--Wigner contraction
\cite{IW,Weimar}.

Some of the quaternionic unitary  CK algebras  are isomorphic; for
instance,  the isomorphism 
\be
\sq_{\k_1,\k_2,\dots,\k_{N-1},\k_N}(N+1)\simeq
\sq_{\k_N,\k_{N-1},\dots,\k_2,\k_1}(N+1)   
\label{na}
\ee
(that interchanges $\k_{ab}\leftrightarrow
\k_{N-b,N-a}$) is provided by the map
\be
\begin{array}{ll}
\j_{ab}\to \j'_{ab}=-\j_{N-b,N-a}&\cr  
\m^1_{ab}\to \m'^1_{ab}=-\m^2_{N-b,N-a}  &\qquad
\e^1_{a}\to \e'^1_{a}=- \e^2_{N-a} \cr
\m^2_{ab}\to \m'^2_{ab}=-\m^1_{N-b,N-a}  &\qquad
\e^2_{a}\to \e'^2_{a}=- \e^1_{N-a} \cr
\m^3_{ab}\to \m'^3_{ab}=-\m^3_{N-b,N-a}  &\qquad
\e^3_{a}\to \e'^3_{a}=- \e^3_{N-a} . 
\end{array}
\label{nb}
\ee

Each algebra in the family of quaternionic unitary  CK
algebras has many
subalgebras isomorphic to orthogonal, unitary, or special unitary CK
algebras, as well as many subalgebras isomorphic to  quaternionic
unitary  algebras in the family
$\sq_\k(M+1)$ with
$M<N$. A clear way to describe this is to denote by
$X_{ab}$ the four generators
$\{\j_{ab},\m^\a_{ab}\}$ 
$(\a=1,2,3)$, by $\e_a$ the set of three generators $\e_a^\a$,  and
arrange the basis generators of 
$\sq_{\k}(N+1)$ as follows:

\begin{center}
\begin{tabular}{ccccc|ccccc}
$\e_0$& $\omm_{01} $&$ \omm_{02} $&$\ldots$&$\omm_{0\, a-1} $&
   $\omm_{0a}$&$\omm_{0\, a+1}$& $\ldots$&$\omm_{0N}$\\
  &$\e_1$& $ \omm_{12} $&$\ldots$&$\omm_{1\, a-1} $&
   $\omm_{1a}$&$\omm_{1\, a+1}$ &$\ldots$&$\omm_{1N}$\\
&   &$\ddots$&$ $&$\vdots$&
   $\vdots$&$\vdots$& &$\vdots$\\
& &$ $&$\e_{a-2}$&$\omm_{a-2\,a-1}$& 
$\omm_{a-2\,a}$&$\omm_{a-2\,a+1}$ &$\ldots$&$\omm_{a-2\,N}$\\
& &$ $&   &$\e_{a-1}$&  $\omm_{a-1\,a}$&$\omm_{a-1\,a+1}$&
 $\ldots$&$\omm_{a-1\,N}$\\
\cline{6-9}
& & &\multicolumn{2}{c}{\,}  
  &$\e_{a}$&$\omm_{a\,a+1}$&
$\ldots$&$\omm_{a  N}$\\ 
& & &\multicolumn{2}{c}{\,}  &   &
  $\ddots$& $$& $\vdots$\\
& & &\multicolumn{2}{c}{\,}  &&
   &  $\e_{N-1}$&$\omm_{N-1\,N}$\\
& & &\multicolumn{2}{c}{\,}  &  &  && $\e_{N}$\\
\end{tabular}
\end{center} 

A Cartan subalgebra is made up of the $N+1$ generators
$\e_{0}^3, \e_{1}^3, \dots, \e_{N}^3$ (in the outermost
diagonal). In this arrangement the generators to the left and
below the rectangle span
subalgebras $\sq_{\k_1,\dots,\k_{a-1}}(a)$ and
$\sq_{\k_{a+1},\dots,\k_N}(N+1-a) $ respectively, while the
generators inside the rectangle do not span a subalgebra unless
$\k_a=0$ (and in this case this is an abelian subalgebra). The
unitary subalgebras $u_\k^\a(N+1) $ appear in a similar way by keeping
only
$J_{ab}$, a single $M_{ab}^\alpha$ out of each $X_{ab}$ and a
single $\e_a^\a$ out of each set $\e_a$ (for a fixed $\alpha$).  By
keeping only
$J_{ab}$ we get the $so_\k(N+1)$ subalgebra.   

If a coefficient $\k_a=0$, then the contracted  algebra has a
semidirect structure
\be
\sq_{\k_1,\dots,\k_{a-1},\k_a=0,\k_{a+1},\dots,\k_N}(N+1)
\equiv  t \odot 
( \sq_{\k_1,\dots,\k_{a-1}}(a)
\oplus
\sq_{\k_{a+1},\dots,\k_N}(N+1-a)) 
\label{ma}
\ee
where $t$ is spanned 
by the generators inside the rectangle (it is an abelian subalgebra of
dimension $4a(N+1-a)$), while
$\sq_{\k_1,\dots,\k_{a-1}}(a)$ and $\sq_{\k_{a+1},\dots,\k_N}(N+1-a)$
are two quaternionic unitary  CK subalgebras spanned by the
generators in the triangles to the left and below the rectangle. When
there are several coefficients $\k_a=0$ the contracted algebra has  
simultaneously several semidirect structures (\ref{ma}).

Notice that when $\k_1=0$ the contracted algebra has the structure
\be
\sq_{0,\k_2,\dots,\k_N}(N+1)
\equiv  t_{4N} \odot
( \sq(1)
\oplus   
\sq_{\k_{2},\dots,\k_N}(N)) 
\label{mb}
\ee
and here the subindex $4N$ in $t$ is the real dimension of the
flat homogeneous space (\ref{ch}) which can be identified with
$\He^N$ endowed with a flat metric given, over $\He$,  by  the
diagonal matrix
$(1, \k_2, \k_2\k_3, \dots, \k_2\k_3\cdots\k_N)$;
when  all these are positive this Lie algebra can be called
inhomogeneous quaternionic unitary  algebra $i\sq(N)$.


\sect{Central extensions}

After having described the structure of the  quaternionic unitary  CK
algebras, we now turn to the second goal of this paper: to give a
complete description of all possible central extensions of the
algebras in the quaternionic unitary  CK family. The outcome of this
study is simple to state: in any dimension, and for all quaternionic
unitary  CK algebras --no matter of how many $\k_a$ are equal or
different from zero--, there are no non-trivial central extensions. 

For any $r$-dimensional Lie algebra with generators
$\{X_1,\dots,X_r\}$ and structure constants
$C_{ij}^k$, a generic central extension by the one-dimensional algebra
generated by
 $\Xi$ will have $(r+1)$ generators
$( X_i,\Xi) $ with commutation relations given by
\be
[X_i,X_j]=\sum_{k=1}^r C_{ij}^k X_k  + \xi_{ij} \Xi  \qquad
[\Xi,X_i]=0 .
\label{de}
\ee
The  extension coefficients or  central charges
$\xi_{ij}$  must be antisymmetric in the indices $i,j$,
$\xi_{ji}=-\xi_{ij}$ and must fulfil the following conditions
coming from the Jacobi identities for the generators  $X_i, X_j, X_l$
in the extended Lie algebra:
\be
\sum_{k=1}^r
\left(
C_{ij}^{k}\xi_{kl}+C_{jl}^{k}\xi_{ki}+C_{li}^{k}\xi_{kj}
\right) =0 .
\label{df}
\ee

If for a set of arbitrary real numbers $\mu_i$ we perform a change
of generators:
\be
X_i\to X'_i=X_i+\mu_i\Xi,
\label{ChangeGens}
\ee
the commutation rules
for the generators $\{X'_i\}$ are given by the expressions
(\ref{de}) with a new set of $\xi_{ij}' = \xi_{ij} -\sum_{k=1}^r
C_{ij}^k \mu_k $, where $\delta\mu(X_i, X_j) = \sum_{k=1}^r
C_{ij}^k \mu_k$ is the two-coboundary generated by $\mu$. Extension
coefficients differing by a two-coboundary correspond to
equivalent extensions; and those extension coefficients which are
a two-coboundary $\xi_{ij}= -\sum_{k=1}^r C_{ij}^k \mu_k $
correspond to trivial extensions; the classes of equivalence of
non-trivial two-cocycles determine the second cohomology group of
the Lie algebra. 


\subsect{Central extensions of the unitary CK subalgebras}

In order to  simplify further computations,  we  first 
state the result about the structure of the central extensions of the
unitary CK  algebra
$u_\k(N+1)$\cite{unitario}, which will naturally appear when studying
the extensions of the quaternionic unitary  CK algebras,  
because each
$\sq_\k(N+1)$ contains three such unitary CK subalgebras.

\medskip

\noindent
{\bf Theorem 3.1.}

\noindent
The  commutation relations of any  central extension
$\overline{u}_\k^\a (N+1)$ of the unitary CK algebra ${u}_\k^\a (N+1)$ 
with generators $\{\j_{ab},\m^\a_{ab},\e^\a_{a}\}$ ($a,b=0,1,\dots,N$
and   quaternionic index $\a$   fixed) by the  one-dimensional
algebra generated by $\Xi$ are 
\be
\begin{array}{ll}
[\j_{ab},\j_{ac}] =\k_{ab}(\j_{bc}+\cextm_{bc}^\a\Xi) &\qquad
[\m_{ab}^\a,\m_{ac}^\a] =\k_{ab}(\j_{bc}+\cextm_{bc}^\a\Xi)\cr
[\j_{ab},\j_{bc}] =-(\j_{ac} +\cextm_{ac}^\a\Xi) &\qquad
[\m_{ab}^\a,\m_{bc}^\a] =\j_{ac}+\cextm_{ac}^\a\Xi\cr
[\j_{ac},\j_{bc}] =\k_{bc}(\j_{ab} + \cextm_{ab}^\a\Xi) &\qquad
[\m_{ac}^\a,\m_{bc}^\a] =\k_{bc}(\j_{ab} + \cextm_{ab}^\a\Xi) \cr
[\j_{ab},\j_{de}]=0 &\qquad
[\m_{ab}^\a,\m_{de}^\a] =0\cr
[\j_{ab},\m_{ac}^\a] =\k_{ab}(\m_{bc}^\a+\cextj_{bc}^\a \Xi) &\qquad
[\m_{ab}^\a,\j_{ac}] =-\k_{ab}(\m_{bc}^\a +  \cextj_{bc}^\a\Xi)
\cr
[\j_{ab},\m_{bc}^\a] =-(\m_{ac}^\a+ \cextj_{ac}^\a \Xi) &\qquad
[\m_{ab}^\a,\j_{bc}] =-(\m_{ac}^\a + \cextj_{ac}^\a \Xi)
\cr
[\j_{ac},\m_{bc}^\a] =-\k_{bc}(\m_{ab}^\a +\cextj_{ab}^\a\Xi) &\qquad
[\m_{ac}^\a,\j_{bc}] =\k_{bc}(\m_{ab}^\a + \cextj_{ab}^\a \Xi)
\cr
\multicolumn{2}{l}{
  [\j_{ab},\e_d^\a] = ( \delta_{ad} -\delta_{bd})(\m_{ab}^\a   
+ \cextj_{ab}^\a \Xi)
\qquad [\j_{ab},\m_{de}^\a] =0}\cr
\multicolumn{2}{l}{
  [\m_{ab}^\a,\e_d^\a] = -( \delta_{ad} -\delta_{bd}) 
(\j_{ab} + \cextm_{ab}^\a\Xi)}
\end{array}
\label{ddaa}
\ee
\be
[\j_{ab},\m^\a_{ab}] =  2\k_{ab}(\e_b^\a - \e_a^\a)+
\cexta_{ab}^\a \Xi
\qquad
[\e_a^\a,\e_b^\a]=\cextb_{a,b}^{\a}\Xi\qquad a< b
\label{da}
\ee
where 
\be
 \cexta_{ab}^\a =\sum_{s=a+1}^b \k_{a\,s-1}\k_{sb}\cexta_{s-1\,s}^\a.
\label{db}
\ee
The extension is characterized by the following types of extension
coefficients:

\noindent
{\bf Type I:} $N(N+1)/2$   coefficients
$\cextj_{ab}^\a$ and  $N(N+1)/2$   coefficients
$\cextm_{ab}^\a$ ($a<b$ and $a,b=0,1,\dots,N$).

\noindent
{\bf Type II:} $N$   coefficients ${\cexta^\a_{a-1\,a}}$
($a=1,\dots,N$).

\noindent
{\bf Type III:} $N(N+1)/2$  coefficients
$\cextb^{\a}_{a,b}$ ($a<b$ and
$a,b=0,1,\dots,N$), satisfying
\be
 \k_{ab} \cextb^{\a}_{a,b}=0\qquad 
\k_{ab} (\cextb^{\a}_{a,c}-\cextb^{\a}_{b,c}) =0\qquad
\k_{bc} (\cextb^{\a}_{a,b}-\cextb^{\a}_{a,c}) =0\qquad a<b<c.
\label{dc}
\ee

\medskip

This theorem expresses the results previously obtained in
\cite{unitario} but in a different basis  (we are using here a
different set of Cartan generators) so that we use another notation
for the extension coefficients.

The extension coefficients are classed into types
according as their behaviour under contraction.
All type I coefficients correspond to central  extensions which are
trivial for all the unitary CK algebras, no matter of how many 
coefficients
$\k_a$ are equal to zero, since they can be  removed at once by means
of the redefinitions
\be
\j_{ab}\to \j_{ab} + \cextm_{ab}^\a\Xi
\qquad
\m_{ab}^\a\to  \m_{ab}^\a + \cextj_{ab}^\a \Xi .
\label{ddcc}
\ee
Each  type II coefficient
${\cexta^\a_{a-1\,a}}$  gives rise to a non-trivial extension if
$\k_a= 0$ and to a trivial one otherwise. That is, these  extensions
become non-trivial through the contraction and they behave as 
pseudoextensions \cite{Aldaya,Azcarraga}. On the other hand, when a
type III coefficient  $\cextb^{\a}_{a,b}$ is non-zero, the extension
that it determines is always non-trivial so that it cannot appear
through a pseudoextension process.
Therefore, the only extensions which can be non-trivial for
a given algebra in the CK family $\overline{u}_\k(N+1)$ are those
appearing in the Lie brackets (\ref{da}).

We also recall that the dimension of the second  cohomology group of
a  unitary CK algebra ${u}_\k(N+1)$ with $n$ coefficients $\k_a$
equal to zero is
\be
{\mbox{dim}}\,
(H^2({u}_\k(N+1),\Ree)=n + \frac {n(n+1)}{2}  =
  \frac {n(n+3)}2 
\label{dd}
\ee
where the first term $n$  
corresponds to the extension coefficients  
${\cexta^\a_{a-1\,a}}$ and the second term $\frac
{n(n+1)}{2}$ to the extensions determined by 
$\cextb^{\a}_{a,b}$.


\subsect{Central extensions of the quaternionic  unitary  CK
algebras}

In the sequel we determine the non-trivial extension
coefficients $\xi_{ij}$ for a  generic centrally
extended quaternionic unitary  CK  algebra $\overline{\sq}_\k(N+1)$
(\ref{de})  by solving the Jacobi identities (\ref{df}).

First, we consider a generic extended unitary CK subalgebra, say
$\overline{u}^1_\k(N+1)$,  spanned by the generators  
$\{\j_{ab},\m^1_{ab},\e^1_{a},\Xi\}$ ($a,b=0,1,\dots,N;\, a<b$) 
with pure quaternionic index equal to 1.  It
is clear that the set of Jacobi identities involving only these  
generators lead to  the results given in the theorem 3.1. Hence, we
find the commutation relations (\ref{ddaa}) and (\ref{da}) with 
extension coefficients denoted
 $\cextj^1_{ab}$, $\cextm^1_{ab}$, ${\cexta^1_{ab}}$ and
$\cextb^1_{a,b}$; we apply the redefinitions (cf.\ (\ref{ddcc}))  
\be
\j_{ab}\to \j_{ab} + \cextm^1_{ab}\Xi
\qquad
\m_{ab}^1\to  \m_{ab}^1 + \cextj_{ab}^1 \Xi 
\label{ka}
\ee
and the Lie brackets of $\overline{u}^1_\k(N+1)\subset
\overline{\sq}_\k(N+1)$ turn out to be
\be
\begin{array}{lll}
[\j_{ab},\j_{ac}] =  \k_{ab}\j_{bc} &\qquad
[\j_{ab},\j_{bc}] =-\j_{ac} &\qquad
[\j_{ac},\j_{bc}] = \k_{bc}\j_{ab}\cr
[\m_{ab}^1,\m_{ac}^1] =\k_{ab} \j_{bc} &\qquad
[\m_{ab}^1,\m_{bc}^1] = \j_{ac} &\qquad
[\m_{ac}^1,\m_{bc}^1] =\k_{bc} \j_{ab} \cr
[\j_{ab},\m_{ac}^1] = \k_{ab}\m_{bc}^1 &\qquad
[\j_{ab},\m_{bc}^1] =-\m_{ac}^1 &\qquad
[\j_{ac},\m_{bc}^1] =-\k_{bc}\m_{ab}^1 \cr
[\m_{ab}^1,\j_{ac}] =-\k_{ab}\m_{bc}^1 &\qquad
[\m_{ab}^1,\j_{bc}] =-\m_{ac}^1 &\qquad
[\m_{ac}^1,\j_{bc}] = \k_{bc}\m_{ab}^1 \cr
[\j_{ab},\j_{de}]= 0 &\qquad
[\m_{ab}^1,\m_{de}^1] =0 &\qquad
[\j_{ab},\m_{de}^1] =0 \cr
\multicolumn{3}{l}{
  [\j_{ab},\e_d^1] = ( \delta_{ad} -\delta_{bd})\m_{ab}^1   
\qquad
  [\m_{ab}^1,\e_d^1] = -( \delta_{ad} -\delta_{bd}) \j_{ab}} 
\end{array}
\label{kb}
\ee
\be 
 [\j_{ab},\m_{ab}^1] = 2\k_{ab}(\e_{b}^1-\e_{a}^1)+
\cexta_{ab}^1 \Xi  
\qquad
  [\e_{a}^1,\e_b^1] = \cextb_{a,b}^{1}\Xi\qquad a< b .
\label{kc}
\ee

The  two remaining extended unitary CK subalgebras 
$\overline{u}^\lambda_\k(N+1)\subset
\overline{\sq}_\k(N+~1)$ with $\lambda=2,3$  are
generated by  $\{\j_{ab}, \m^\lambda_{ab},\e^\lambda_{a},\Xi\}$
(hereafter we shall reserve $\lambda$ to stand exclusively for the
quaternionic indices $\lambda=2,3$, whereas  
$\a,\b,\c$ are allowed to take on any value $1,2,3$). The subalgebras
$\overline{u}^\lambda_\k(N+1)$ have generic extended   Lie
brackets (as (\ref{de})) except for the common orthogonal CK
subalgebra
${so}_\k(N+1)$ spanned by the generators
$\{ \j_{ab}\}$ which is non-extended and whose Lie brackets are 
already written in (\ref{kb}). For the two remaining unitary
subalgebras, we have already used up the redefinition concerning to
the common generators in ${so}_\k(N+1)$, so we cannot apply
directly the results of the theorem 3.1 and we
have to compute their corresponding Jacobi identities by taking
into account that initially both contain a non-extended 
${so}_\k(N+1)$. As the  equations so obtained are similar to those
written in detail   in \cite{unitario}   we omit them  and give the
final result.  The extension coefficients that appear are denoted
 $\cextj^\lambda_{ab}$, $\cextm^\lambda_{a\,a+1}$,
${\cexta^\lambda_{ab}}$ and $\cextb^\lambda_{a,b}$, for
$\lambda=2,3$; the Lie brackets of  
$\overline{u}^\lambda_\k(N+1)$ read
\be
\begin{array}{ll}
\multicolumn{2}{l}{[\m_{ab}^\lambda ,\m_{ac}^\lambda ] =\k_{ab}
\j_{bc}  \qquad
[\m_{ab}^\lambda ,\m_{bc}^\lambda ] =\j_{ac}  \qquad
[\m_{ac}^\lambda ,\m_{bc}^\lambda ] =\k_{bc} \j_{ab} } \cr
[\j_{ab},\m_{ac}^\lambda ] =\k_{ab}(\m_{bc}^\lambda
+\cextj_{bc}^\lambda 
\Xi) 
&\qquad [\m_{ab}^\lambda ,\j_{ac}] =-\k_{ab}(\m_{bc}^\lambda  + 
\cextj_{bc}^\lambda \Xi)
\cr
[\j_{ab},\m_{bc}^\lambda ] =-(\m_{ac}^\lambda +
\cextj_{ac}^\lambda  \Xi) 
&\qquad [\m_{ab}^\lambda ,\j_{bc}] =-(\m_{ac}^\lambda  +
\cextj_{ac}^\lambda  \Xi)
\cr
[\j_{ac},\m_{bc}^\lambda ] 
=-\k_{bc}(\m_{ab}^\lambda  +\cextj_{ab}^\lambda \Xi)
 &\qquad [\m_{ac}^\lambda ,\j_{bc}] =\k_{bc}(\m_{ab}^\lambda  +
\cextj_{ab}^\lambda  \Xi)\cr
\multicolumn{2}{l}{[\j_{ab},\m_{de}^\lambda ] =0
  \qquad  \qquad
[\m_{ab}^\lambda ,\m_{de}^\lambda ] =0}\cr
\multicolumn{2}{l}{
  [\j_{ab},\e_d^\lambda ] = ( \delta_{ad}
-\delta_{bd})(\m_{ab}^\lambda     + \cextj_{ab}^\lambda  \Xi) }\cr
\multicolumn{2}{l}{
  [\m_{ab}^\lambda ,\e_d^\lambda ] = -( \delta_{ad} -\delta_{bd})
\j_{ab}  \qquad b>a+1}\cr
\multicolumn{2}{l}{
  [\m_{a\, a+1}^\lambda,\e_d^\lambda ] =  
-( \delta_{ad} -\delta_{a+1\, d})( \j_{a\, a+1}  + \cextm_{a\,
a+1}^\lambda 
\Xi )  }\cr
\end{array}
\label{kd}
\ee
\be
[\j_{ab},\m _{ab}^\lambda] =  2\k_{ab}(\e_b^\lambda  - \e_a^\lambda
)+ \cexta_{ab}^\lambda  \Xi
\qquad
[\e_a^\lambda ,\e_b^\lambda ]=\cextb_{a,b}^\lambda\Xi\qquad a< b .
\label{ke}
\ee
The coefficients ${\cexta^\lambda_{ab}}$ and $\cextb^\lambda_{a,b}$
($\lambda=2,3$) are characterized by the theorem 3.1 (see
(\ref{db}) and (\ref{dc})),  while the extensions
$\cextm^\lambda_{a\,a+1}$ are subjected to the relations
\be
\k_a\cextm^\lambda_{a\,a+1}=0\qquad 
\k_{a+2}\cextm^\lambda_{a\,a+1}=0 .
\label{kf}
\ee
Notice that now the coefficients $\cextm^\lambda_{ab}$ with $b>a+1$
are zero (this is a direct consequence of the presence of the 
non-extended  ${so}_\k(N+1)$). We now apply  the redefinitions
given by
\be
\m_{ab}^\lambda\to  \m_{ab}^\lambda + \cextj_{ab}^\lambda \Xi\qquad
\lambda= 2,3  
\label{kg}
\ee
and a glance to (\ref{kd}) shows that the corresponding extensions
are always trivial, so the extension coefficients
$\cextj_{ab}^\lambda$ are eliminated. 

At this point the complete set of Lie brackets of
$\overline{\sq}_\k(N+1)$ turn out to be 
\be
\begin{array}{lll}
[\j_{ab},\j_{ac}] =  \k_{ab}\j_{bc} &\qquad
[\j_{ab},\j_{bc}] =-\j_{ac} &\qquad
[\j_{ac},\j_{bc}] = \k_{bc}\j_{ab}\cr
[\m_{ab}^\a,\m_{ac}^\a] =\k_{ab} \j_{bc} &\qquad
[\m_{ab}^\a,\m_{bc}^\a] = \j_{ac} &\qquad
[\m_{ac}^\a,\m_{bc}^\a] =\k_{bc} \j_{ab} \cr
[\j_{ab},\m_{ac}^\a] = \k_{ab}\m_{bc}^\a &\qquad
[\j_{ab},\m_{bc}^\a] =-\m_{ac}^\a &\qquad
[\j_{ac},\m_{bc}^\a] =-\k_{bc}\m_{ab}^\a \cr
[\m_{ab}^\a,\j_{ac}] =-\k_{ab}\m_{bc}^\a &\qquad
[\m_{ab}^\a,\j_{bc}] =-\m_{ac}^\a &\qquad
[\m_{ac}^\a,\j_{bc}] = \k_{bc}\m_{ab}^\a \cr
[\j_{ab},\j_{de}]= 0 &\qquad
[\m_{ab}^\a,\m_{de}^\a] =0 &\qquad
[\j_{ab},\m_{de}^\a] =0 \cr
\multicolumn{3}{l}{
  [\j_{ab},\e_d^\a] = ( \delta_{ad} -\delta_{bd})\m_{ab}^\a   
\qquad
  [\m_{ab}^1,\e_d^1] = -( \delta_{ad} -\delta_{bd}) \j_{ab}}\cr
\multicolumn{3}{l}{
  [\m_{ab}^\lambda,\e_d^\lambda] = -( \delta_{ad} -\delta_{bd})
\j_{ab}\qquad b>a+1}
\end{array}
\label{dg}
\ee
\be
  [\m_{a\, a+1}^\lambda,\e_d^\lambda ] =  
-( \delta_{ad} -\delta_{a+1\, d})( \j_{a\, a+1}  + \cextm_{a\,
a+1}^\lambda 
\Xi ) 
\label{ddgg}
\ee
\be 
 [\j_{ab},\m_{ab}^\a] = 2\k_{ab}(\e_{b}^\a-\e_{a}^\a)+
\cexta_{ab}^\a \Xi  
\qquad
  [\e_{a}^\a,\e_b^\a] = \cextb_{a,b}^{\a}\Xi\qquad a< b
\label{dh}
\ee
\be
\begin{array}{l}
[\m_{ab}^\a,\m_{ac}^\b] = \k_{ab}\ep_{\a\b\c}\m_{bc}^\c + 
\ep_{\a\b\c}\mm_{ab,ac}^{\a,\b}\Xi \cr
[\m_{ab}^\a,\m_{bc}^\b] = \ep_{\a\b\c}\m_{ac}^\c + 
\ep_{\a\b\c}\mm_{ab,bc}^{\a,\b}\Xi  \cr
[\m_{ac}^\a,\m_{bc}^\b] = \k_{bc}\ep_{\a\b\c}\m_{ab}^\c+ 
\ep_{\a\b\c}\mm_{ac,bc}^{\a,\b}\Xi  \cr
[\m_{ab}^\a,\m_{de}^\b] =  
\ep_{\a\b\c}\mm_{ab,de}^{\a,\b}\Xi \cr
[\m_{ab}^\a,\m_{ab}^\b] = 2\k_{ab}\ep_{\a\b\c}(\e_a^\c + \e_b^\c)+ 
\ep_{\a\b\c}\mm_{ab}^{\a,\b}\Xi  \cr
[\m_{ab}^\a,\e_{d}^\b] =(\delta_{ad}
+\delta_{bd})\ep_{\a\b\c}\m_{ab}^\c+ 
\ep_{\a\b\c}\me_{ab,d}^{\a,\b}\Xi  \cr
[\e_{a}^\a,\e_{b}^\b] =2\delta_{ab}\ep_{\a\b\c}\e_{a}^\c + 
\ep_{\a\b\c}\cextb_{a,b}^{\a,\b}\Xi   .
\end{array}
\label{di}
\ee
Therefore the  Lie brackets (\ref{dg}) 
are non-extended, the extension coefficients  $\cextm_{a\,
a+1}^\lambda$ appearing in (\ref{ddgg}) satisfy (\ref{kf}), the  
coefficients  of the commutators  (\ref{dh}) are characterized by
the theorem 3.1, and  the extension coefficients  in  the
commutators (\ref{di}) are still generic, the redefinitions
(\ref{ka}) and (\ref{kg}) having been already incorporated in the
brackets (\ref{di}).

The list of all remaining extension coefficients is
\be
\cextm_{a\, a+1}^\lambda \qquad \cexta_{ab}^\a\qquad
 \cextb_{a,b}^{\a}\qquad
\mm_{ab,de}^{\a,\b}\qquad
\mm_{ab}^{\a,\b}\qquad
\me_{ab,d}^{\a,\b}\qquad
\cextb_{a,b}^{\a,\b}  
\label{dj}
\ee
where the two quaternionic indices $\a$, $\b$ are always different.
We sort the coefficients  $\mm_{ab,de}^{\a,\b}$,
$\me_{ab,d}^{\a,\b}$ and
$\cextb_{a,b}^{\a,\b}$  into several subsets as follows:

\noindent
$\bullet$ Coefficients  $\mm_{ab,de}^{\a,\b}$ involving \emph{four}
different indices $a,b,d,e$. If we rename  and sort these four
different indices as
$a<b<c<d$ these coefficients are
\be
\mm_{ab,cd}^{\a,\b}\qquad 
\mm_{ac,bd}^{\a,\b}\qquad \mm_{ad,bc}^{\a,\b} .
\label{dk}
\ee

\noindent
$\bullet$ Coefficients $\mm_{ab,de}^{\a,\b}$  involving \emph{three}
different indices. If we write the indices as $a<b<c$ the
coefficients are
\be
\mm_{ab,ac}^{\a,\b}\qquad
 \mm_{ab,bc}^{\a,\b}\qquad \mm_{ac,bc}^{\a,\b} .
\label{dl}
\ee

\noindent
$\bullet$ Coefficients $\me_{ab,d}^{\a,\b}$ with \emph{two} different
indices $a<b$ and a third one $d\in \{a,b\}$:
\be
\me_{ab,a}^{\a,\b}\qquad \me_{ab,b}^{\a,\b} .
\label{dm}
\ee

\noindent
$\bullet$ Coefficients $\me_{ab,d}^{\a,\b}$ with \emph{two} different
indices $a<b$ and a third index $d\notin \{a,b\}$:
\be
\me_{ab,d}^{\a,\b}  .
\label{dmm}
\ee

\noindent
$\bullet$ Coefficients $\cextb_{a,b}^{\a,\b}$ with  \emph{two}
different indices $a<b$:
\be
\cextb_{a,b}^{\a,\b}  .
\label{dn}
\ee

\noindent
$\bullet$ Coefficients $\cextb_{a,b}^{\a,\b}$ with a \emph{single}  
index $a$:
\be
\cextb_{a,a}^{\a,\b} .
\label{do}
\ee

In the sequel we proceed to   compute the  Jacobi identities involving
the above coefficients;  the results
obtained in any equation will be automatically introduced in
any further equation, so the order we consider for enforcing the
Jacobi identities is an integral part of the derivation, and should be
respected. We denote the  Jacobi identity (\ref{df}) of the generators 
$X_i$, $X_j$ and $X_l$ by $\{X_i, X_j ,X_l\}$.

The following equations imply the vanishing of   some
 coefficients:
\be
\begin{array}{ll}
 \{\m_{a\, a+1}^3,\e_{a}^1,\e_{a+1}^2\}: &\quad
\cextm_{a\, a+1}^2=0 \cr
 \{\m_{a\, a+1}^2,\e_{a}^1,\e_{a+1}^3\}: &\quad
\cextm_{a\, a+1}^3=0
\end{array}
\label{eebb}
\ee
\be
\{\e_{a}^\c,\e_{a}^\b,\e_{b}^\a\}:\quad  \cextb_{a,b}^{\a}=0 
\label{ea}
\ee
\be
\begin{array}{ll}
 \{\m_{ab}^\a,\m_{ac}^\a,\e_{c}^\c\}: &\quad \mm_{ab,ac}^{\a,\b}=0\cr
\{\m_{ab}^\a,\m_{bc}^\a,\e_{c}^\c\}: &\quad \mm_{ab,bc}^{\a,\b}=0\cr
\{\m_{ac}^\a,\m_{bc}^\a,\e_{b}^\c\}: &\quad \mm_{ac,bc}^{\a,\b}=0  
\end{array}
\label{eb}
\ee
\be
\begin{array}{ll}
 \{\j_{ab} ,\m_{cd}^\b,\e_{b}^\a\}: &\quad \mm_{ab,cd}^{\a,\b}=0\cr
\{\j_{bc} ,\m_{ad}^\a,\e_{c}^\b\}: &\quad \mm_{ad,bc}^{\a,\b}=0\cr
\{\j_{ab} ,\m_{bc}^\a,\m_{bd}^\b\}: &\quad
\mm_{ac,bd}^{\a,\b}-\mm_{ad,bc}^{\b,\a}=0 
\end{array}
\label{ec}
\ee
\be
\begin{array}{ll}
 \{\m_{ab}^\b,\e_{a}^\b,\e_{b}^\c\}: &\quad \me_{ab,a}^{\a,\b}=0\cr
\{\m_{ab}^\b,\e_{b}^\b,\e_{a}^\c\}: &\quad \me_{ab,b}^{\a,\b}=0 
\end{array}
\label{ed}
\ee
\be
\{\m_{ab}^\c,\e_{a}^\b,\e_{d}^\b\}:  \qquad \me_{ab,d}^{\a,\b}=0 
\label{ee}
\ee
\be
\{\e_{a}^\a,\e_{b}^\a,\e_{b}^\c\}:  \qquad \cextb_{a,b}^{\a,\b}=0 
\label{ef}
\ee
so that the only remaining  coefficients are
$\cexta_{ab}^\a$, 
$\mm_{ab}^{\a,\b}$ and $\cextb_{a,a}^{\a,\b}$. The Jacobi identities
\be
\begin{array}{ll}
 \{\j_{ab},\m_{ab}^\a,\e_{a}^\b\}: &\quad 2\k_{ab}\cextb_{a,a}^{\a,\b}
-\mm_{ab}^{\a,\b}+\cexta_{ab}^\c =0\cr
\{\j_{ab},\m_{ab}^\a,\e_{b}^\b\}: &\quad 2\k_{ab}\cextb_{b,b}^{\a,\b}
-\mm_{ab}^{\a,\b}-\cexta_{ab}^\c =0 
\end{array}
\label{eg}
\ee
allows us to express the coefficients $\cexta_{ab}^\a$, 
$\mm_{ab}^{\a,\b}$ in terms of the $\cextb_{a,a}^{\a,\b}$ as follows
\be
\begin{array}{l}
\cexta_{ab}^\c=\k_{ab}(\cextb_{b,b}^{\a,\b}-\cextb_{a,a}^{\a,\b})\cr
\mm_{ab}^{\a,\b}=\k_{ab}(\cextb_{b,b}^{\a,\b}+\cextb_{a,a}^{\a,\b}) .
\end{array}
\label{eh}
\ee
Notice that the first equation is consistent  with the
relation (\ref{db}). Hence  the only Lie brackets of
$\overline{\sq}_\k(N+1)$ (\ref{dg})--(\ref{di}) which still involve
extension coefficients are
\be
\begin{array}{l}
 [\j_{ab},\m_{ab}^\c] = 2\k_{ab}\left\{ (\e_{b}^\c + \frac 12 
\cextb_{b,b}^{\a,\b} \Xi) -  (\e_{a}^\c + \frac 12
\cextb_{a,a}^{\a,\b} \Xi)\right\}  \cr 
[\m_{ab}^\a,\m_{ab}^\b] = 2\k_{ab}\ep_{\a\b\c}\left\{
(\e_a^\c +\frac 12 \cextb_{a,a}^{\a,\b} \Xi) +( \e_b^\c+ 
\frac 12 \cextb_{b,b}^{\a,\b} \Xi)\right\} \cr
[\e_{a}^\a,\e_{a}^\b] =2\ep_{\a\b\c}(\e_{a}^\c + 
\frac 12 \cextb_{a,a}^{\a,\b} \Xi ).
\end{array}
\label{ei}
\ee
These equations clearly suggest to introduce  the redefinition
\be 
\e_{a}^\c \to \e_{a}^\c + 
\frac 12 \cextb_{a,a}^{\a,\b} \Xi 
\label{ej}
\ee
which explicitly shows the triviality of all the  extensions
determined by the  coefficients 
$\cextb_{a,a}^{\a,\b}$ (and consequently, by all the
$\cexta_{ab}^\a$  and $\mm_{ab}^{\a,\b}$). Therefore it is
not necessary to compute more Jacobi  identities  and  we can
conclude that  the most general central extension
$\overline{\sq}_\k(N+1)$ of any algebra in this family is always
trivial. 
 
This result can be summed up in the  following
statement:

\medskip

\noindent
{\bf Theorem 3.2.}

\noindent
The second cohomology group $H^2({\sq}_\k(N+1),\Ree)$ of  any Lie
algebra belonging to the quaternionic unitary  CK family is always
trivial, for any  $N$ and for any values of the set of constants
$\k_1,
\k_2, \dots, \k_N$:
\be
{\mbox{dim}}\,
(H^2({\sq}_\k(N+1),\Ree) )=0 .
\label{ek}
\ee


\sect{Concluding remarks}

This paper completes the study of cohomology of the quasi-simple
or CK Lie algebras in the three main series (orthogonal, unitary and
quaternionic unitary), as associated to antihermitian matrices over
$\Ree, \Cee$ or $\He$. In contrast to the quasi-orthogonal or 
quasi-unitary cases, where the dimension of the second  cohomology
group of a generic algebra in the CK family ranges between $0$ for
the simple algebras and a maximum positive value for the most
contracted algebra (with all $\k_a=0$), all the central extensions of
any of the algebras in the quaternionic quasi-unitary family are
always trivial, even for the most contracted algebra. Therefore from
the three types of extensions found in the quasi-orthogonal or 
quasi-unitary cases, only the first type (extensions which are
trivial for all the algebras in the family) is present here. However
we should remark the suitability of a CK approach to the study of the
central extensions of a complete family, because a case-by-case study
(for any given algebra in the family) would be not more easy than the
general analysis we have performed. 

In addition to these three \emph{main} families of CK algebras,
whose simple members $so(p, q), su(p, q), \sq(p,q)$ can be realised
as antihermitian matrices over either $\Ree$, $\Cee$  or $\He$, there
are other CK families. In the $C_{N+1}$ Cartan series, the
remaining real Lie algebra is the real
symplectic $sp(2(N+1), \Ree)$, which can be interpreted in terms of CK
families either as the single simple member of its own CK family 
$sp_{\k_1, \dots, \k_{N}}(2(N+1), \Ree)$, or  alternatively and  more
like the interpretation in this paper, as the unitary family
$u_{\k_1, \dots,
\k_{N}}((N+1), \He')$ over the algebra of  the split quaternions
$\He'$ (a pseudo-orthogonal variant of quaternions, where 
$i_1, i_2, i_3$ still anticommute, but their  squares are $i_1^2=-1,
i_2^2=1, i_3^2=1$; this is not a division algebra).  The
cohomology properties of algebras in this CK family could be
studied using an approach similar to that made in
this paper for the quaternionic unitary  CK algebras.  This study, a
well as the study of the central extensions of the CK series of the
real Lie algebras
 $su^*(2r)\approx sl(r, \He), so^*(2N), sl(N+1, \Ree)\approx su(N+1,
\Cee')$ not included in the three main `signature' series is worth of
a separate consideration.


\bigskip\bigskip

\noindent
{\Large{{\bf Acknowledgments}}}

\bigskip

This  work was partially supported by DGICYT (project  PB94--1115)
from the Ministerio de Educaci\'on y Cultura de Espa\~na   and by
Junta de Castilla y Le\'on (Projects CO1/396 and CO2/297). 


\end{document}